\documentclass[conference]{IEEEtran}
\usepackage{cite}
\usepackage{amsmath,amssymb,amsfonts,amsthm}
\usepackage{graphicx}
\usepackage{epstopdf}
\usepackage{algorithm}
\usepackage{algorithmic}
\usepackage{enumerate}
\usepackage{subfigure}

\newcommand{\algorithmicinput}{\textbf{Input:}}
\newcommand{\INPUT}{\item[\algorithmicinput]}

\newcommand{\algorithmicoutput}{\textbf{Output:}}
\newcommand{\OUTPUT}{\item[\algorithmicoutput]}
\newtheorem{remark}{Remark}
\theoremstyle{plain}

\begin{document}

\title{High-Resolution Channel Estimation for Intelligent Reflecting Surface-Assisted MmWave Communications}

\author{
\IEEEauthorblockN{Chenglu Jia\IEEEauthorrefmark{2}, Junqiang Cheng\IEEEauthorrefmark{2}, Hui Gao\IEEEauthorrefmark{2}, and Wenjun Xu\IEEEauthorrefmark{3}}
\IEEEauthorblockA{\IEEEauthorrefmark{2}Key Lab of Trustworthy Distributed Computing and Service, Ministry of Education}
\IEEEauthorblockA{\IEEEauthorrefmark{3}Key Lab of Universal Wireless Communications, Ministry of Education\\
Beijing University of Posts and Telecommunications, Beijing, China, 100876\\
Email: \{chenglujia, jqcheng, huigao, wjxu\}@bupt.edu.cn}
}
\maketitle

\begin{abstract}
In this paper, we study the high-resolution channel estimation problem for intelligent reflecting surface (IRS)-assisted millimeter wave (mmWave) multiple-input-multiple-output (MIMO) communications, which is a prerequisite to guarantee further high-rate data transmission. Considering the typical sparsity of mmWave channels, we formulate the cascaded channel estimation problem from a sparse signal recovery perspective, and then propose a novel two-step cascaded channel estimation protocol to estimate the cascaded user-IRS-base station channel with high-resolution for IRS-assisted mmWave MIMO communications. More specifically, the first step is to estimate the coarse angular domain information (ADI) and further establish the robust uplink by beam training. In the second step, by exploiting the coarse ADI, an adaptive grid matching pursuit (AGMP) algorithm is proposed to estimate the high-resolution cascaded channel state information (CSI) with low complexity. Simulation results verify that the proposed two-step channel estimation protocol significantly outperforms the state-of-the-art scheme, i.e., beam training based channel estimation, and meanwhile can reap near-optimal system performance achieved by perfect CSI.
\end{abstract}


%
\IEEEpeerreviewmaketitle

\section{Introduction}
%
%
%
%
Millimeter wave (mmWave), as an essential technology for 5G mobile communications, has attracted considerable attentions from both academia and industry thanks to its sufficient unoccupied spectrum resource \cite{6515173}. However, mmWave transmission inherently suffers from severe path-loss due to its high operating frequency in the range of 30-300 GHz, which poses a critical challenge for practical implementation \cite{gao2016energy}. Aiming to compensate the link-budget gap, large-scale antenna arrays are usually equipped at transceivers to achieve directional beamforming with focused energy. However, the high directionality coupled with poor penetrability makes mmWave communications vulnerable to blockage events, which significantly deteriorates the coverage capability and limits its applications in urban cellular systems.

Recently, intelligent reflecting surface (IRS) is emerging as a promising technology to achieve high spectral and energy efficiency, and has been integrated into various communication systems on sub-6G band \cite{a8796365, a8746155, a8741198}. Specifically, IRS, acting as a planar array, is composed of a large number of low-cost passive reflective elements, which can smartly steer the incident signals towards the dedicated directions via a software-controlled manner \cite{8647620}. Most recently, IRS is also considered as an energy-efficient solution to combat the blockage of mmWave communications by creating the \emph{virtual} light-of-sight (LOS) path \cite{wang2019intelligent, cao2019intelligent, wang2019joint}. Specifically, the authors in \cite{wang2019intelligent, cao2019intelligent, wang2019joint} investigated the joint beamforming issue and verified that the presence of IRS can improve the mmWave coverage obviously. However, most of these works are based on the assumption of perfect channel state information (CSI), while CSI acquirement is actually a very challenging task, especially for IRS-assisted mmWave communications. Specifically, on the one hand, the introduce of IRS complicates the communication topology, and the user (UE)-IRS and IRS-base station (BS) channels need to be estimated at the same time. On the other hand, the passive IRS does not possess any signal transmitting/receiving/processsing capabilities due to the lack of active radio frequency (RF) chains, which results in that the cascaded channel can be only estimated at receivers.

There are already some channel estimation schemes for IRS-assisted wireless communications on sub-6GHz band \cite{8879620,mishra2019channel,8937491,you2019intelligent, you2019progressive}. However, these schemes are not applicable to the IRS-assisted mmWave multiple-input-multiple-output (MIMO) communications. The reasons are listed as follows: 1) the channel matrix has a large size due to the large-scale antenna arrays deployed at both transceivers and IRS for mmWave communications, which makes the entry-wise channel estimation schemes suffer from prohibitive computation complexity and pilot overhead; 2) the random phase of IRS is not reasonable for high-directional mmWave links since beam misalignment can significantly degrade the channel estimation performance. Despite these difficulties, the research on channel estimation for IRS-assisted mmWave communications is still very limited \cite{wang2019compressed, wan2020broadband, taha2019enabling, ning2019channel}. Specifically, by exploiting the sparsity of mmWave channels, the authors in \cite{wang2019compressed, wan2020broadband} studied the compressive sensing (CS) based cascaded channel estimation for IRS-assisted multiple-input-single-output (MISO) systems with single-antenna transmitter/receiver. Moreover, the authors in \cite{taha2019enabling} divided the cascaded channel estimation into two stages, i.e., estimating the BS-IRS channel and UE-IRS channel by equipping IRS with active RF chains, which makes channel estimation a high energy-consuming operation. Most recently, the authors in \cite{ning2019channel} proposed a beam training based channel estimation scheme. However, due to the imperfect hardware manufacturing technique, it is impractical to fabricate IRS with infinite phase-resolution, which severely degrades the accuracy of channel estimation. We can observe that there is little discussion on high-resolution cascaded channel estimation for IRS-assisted mmWave MIMO communications, while it is imperative for further system optimization and data transmission.

In this paper, by exploiting the sparsity of mmWave channels, we formulate the cascaded channel estimation into a sparse recovery problem. Taking the hardware-constraint into consideration, we propose a two-step cascaded channel estimation protocol to acquire high-resolution CSI. In the first step, the beam training procedure is utilized to estimate coarse angular domain information (ADI), which is regarded as an important prior information for uplink establishment and further high-resolution channel estimation. In the second step, enhanced by the prior ADI acquired by the first step, a low-complexity adaptive grid matching pursuit (AGMP) algorithm is proposed to estimate the high-resolution cascaded CSI. In particular, the beam training based channel estimation scheme \cite{ning2019channel} is adopted as the benchmark. The simulation results show that the proposed two-step channel estimation protocol provides a more accurate CSI as compared to the benchmark and reaps better system performance in terms of spectral efficiency (SE).

\textbf{Notation}: $j=\sqrt{-1}$. $\mathbf{A}$, $\mathbf{a}$ and $\mathcal{A}$ are matrix, vector and set, respectively. $\mathbf{A}^{H}$, $\mathbf{A}^{*}$ and $\mathbf{A}^{T}$ denote conjugate transpose, conjugate and transpose of $\mathbf{A}$, respectively. $\mathbf{A}(i)$ denotes the $i$th row of $\mathbf{A}$. $\mathbb{E}\left(\cdot\right)$ denotes the expectation operation. $\left\Vert \mathbf{A}\right\Vert _{F}$ is the Frobenius norm of $\mathbf{A}$. $\textrm{vec}\left(\mathbf{A}\right)$ represents the column-wise vectorization of $\mathbf{A}$, and its inverse operation is $\textrm{unvec}\left(\mathbf{a}\right)$. $\mathbf{A}\otimes\mathbf{B}$ denotes the Kronecker product of $\mathbf{A}$ and $\mathbf{B}$.
\section{Channel Model}

\begin{figure}[tp]
  \centering
  \includegraphics[width=0.48\textwidth]{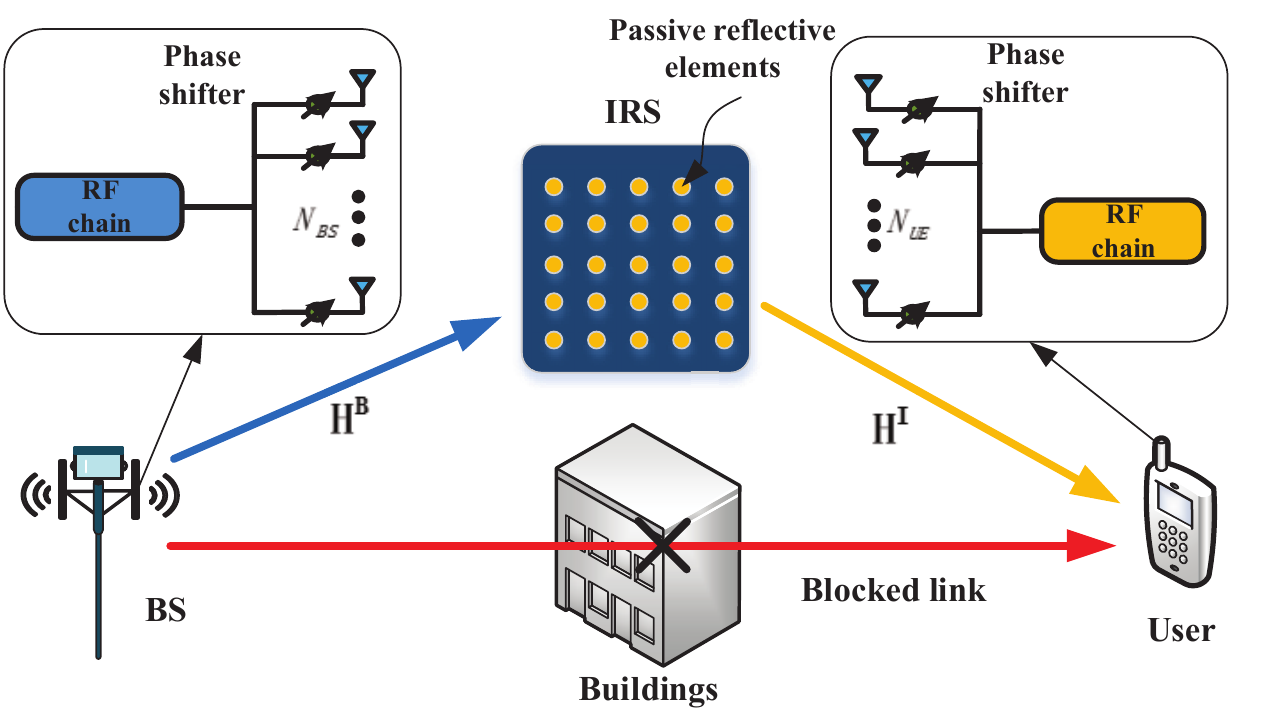}
  \caption{Illustration of the considered IRS-assisted mmWave MIMO system.}
  \label{transceiver}
\end{figure}

 We consider an IRS-assisted mmWave MIMO system containing a single BS, a single IRS and a single user, which is shown in Fig. \ref{transceiver}. More specifically, the IRS is deployed to provide a \emph{virtual} LOS link for the user that is blocked by the obstacles. Similar to \cite{8122055}, we consider that both BS and user are equipped with an analog antenna architecture consisting of a full-connection uniform linear array, where the numbers of antennas are $N_{BS}$ and $N_{UE}$, respectively. The IRS is equipped with a planar array consisting of $N_{I}$ passive reflective elements whose amplitude and phase can be adjusted dynamically via the IRS-controller.

The geometric channel model is adopted to characterize the channels for the considered IRS-assisted mmWave MIMO communications, and the channel matrix $\mathbf{H}\in \mathbb{C}^{N_{R}\times N_{T}}$ is expressed as \cite{7400949}
\begin{equation}\label{channel}
 \mathbf{H}=\sqrt{\frac{N_{T}N_{R}}{L}}\overset{L}{\underset{l=1}{\sum}}\alpha_{l}\mathbf{a}_{R}\left(N_{R},\varPhi_{l}\right)\mathbf{a}_{T}\left(N_{T},\varOmega_{l}\right)^{H},
\end{equation}
where $N_T$ and $N_R$ denotes the number of antennas at transmitter and receiver, respectively. $L$ is the number of path, and $\alpha_{l}$ is the channel complex gain of the $l$th path. $\theta$ and $\varphi$ denote the angle-of-arrival (AoA) and angle-of-departure (AoD). For convenience, we have $\varPhi=\text{cos}\left(\theta\right)$ and $\varOmega=\text{cos}\left(\varphi\right)$, which are called AoA and AoD in the rest of this paper. $\mathbf{a}_{R}\left(\cdot\right)$ and $\mathbf{a}_{T}\left(\cdot\right)$ are the steering vectors at receiver and transmitter, respectively, which are given by
\begin{equation}\label{steer}
\begin{split}
    & \mathbf{a}_{R}\left(N_{R},\varPhi\right)=\frac{1}{\sqrt{N_{R}}}\left[1,e^{j2\pi\frac{d}{\lambda}\varPhi},...,e^{j2\pi\left(N_{R}-1\right)\frac{d}{\lambda}\varPhi}\right]^{T}, \\
    & \mathbf{a}_{T}\left(N_{T},\varOmega\right)=\frac{1}{\sqrt{N_{T}}}\left[1,e^{j2\pi\frac{d}{\lambda}\varOmega},...,e^{j\pi\left(N_{T}-1\right)\frac{d}{\lambda}\varOmega}\right]^{T},
\end{split}
\end{equation}
where $\lambda$ is the wavelength of the carrier, and $d$ is the antenna spacing which is equal to the half-wavelength. The channel model in (\ref{channel}) can be written in a more compact form as

\begin{equation}\label{compact_channel}
  \mathbf{H}=\mathbf{A}_{R}\mathbf{H}_{a}\mathbf{A}_{T}^{H},
\end{equation}
where $\mathbf{H}_{a}=\sqrt{\frac{N_{T}N_{R}}{L}}\text{diag}\left(\alpha_{1},...,\alpha_{L}\right)$ represents the angular domain channel, $\mathbf{A}_{R}=\left[\mathbf{a}_{R}\left(N_{R},\varPhi_{1}\right),...,\mathbf{a}_{R}\left(N_{R},\varPhi_{L}\right)\right]\in\mathbb{C}^{N_{R}\times L}$, $\mathbf{A}_{T}=\left[\mathbf{a}_{T}\left(N_{T},\varOmega_{1}\right),...,\mathbf{a}_{T}\left(N_{T},\varOmega_{L}\right)\right]\in\mathbb{C}^{N_{T}\times L}$.

For the considered IRS-assisted mmWave system, let $\mathbf{H}^{B}\in\mathbb{C}^{N_{I}\times N_{BS}}$ denotes the channel from BS to IRS, $\mathbf{H}^{I}\in\mathbb{C}^{N_{UE}\times N_{I}}$ denotes the channel from IRS to UE. Assuming that the direct link from BS to UE is blocked, we mainly focus on the cascaded channel acquirement. The BS-IRS-UE cascaded channel $\mathbf{H}^{S}\in\mathbb{C}^{N_{UE}\times N_{BS}}$ can be expressed as
\begin{equation}\label{syn_channel}
  \mathbf{H}^{S}=\mathbf{H}^{I}\mathbf{\Theta}\mathbf{H}^{B},
\end{equation}
where $\mathbf{\Theta}=\text{diag}\left(\beta_{1}e^{j\phi_{1}},...,\beta_{N_{I}}e^{j\phi_{N_{I}}}\right)$ denotes the phase-shift matrix of IRS with $\beta_{n}\in\left[0,1\right]$ and $\phi_{n}\in\left[0,2\pi\right)$ being the reflection amplitude and phase-shift parameters of the $n$th passive reflective element, respectively. Without loss of generality, we assume $\beta_{n}=1$ for all elements to maximize beamforming gain brought by IRS \cite{8647620}.

\begin{figure}[tp]
  \centering
  \includegraphics[width=0.4\textwidth]{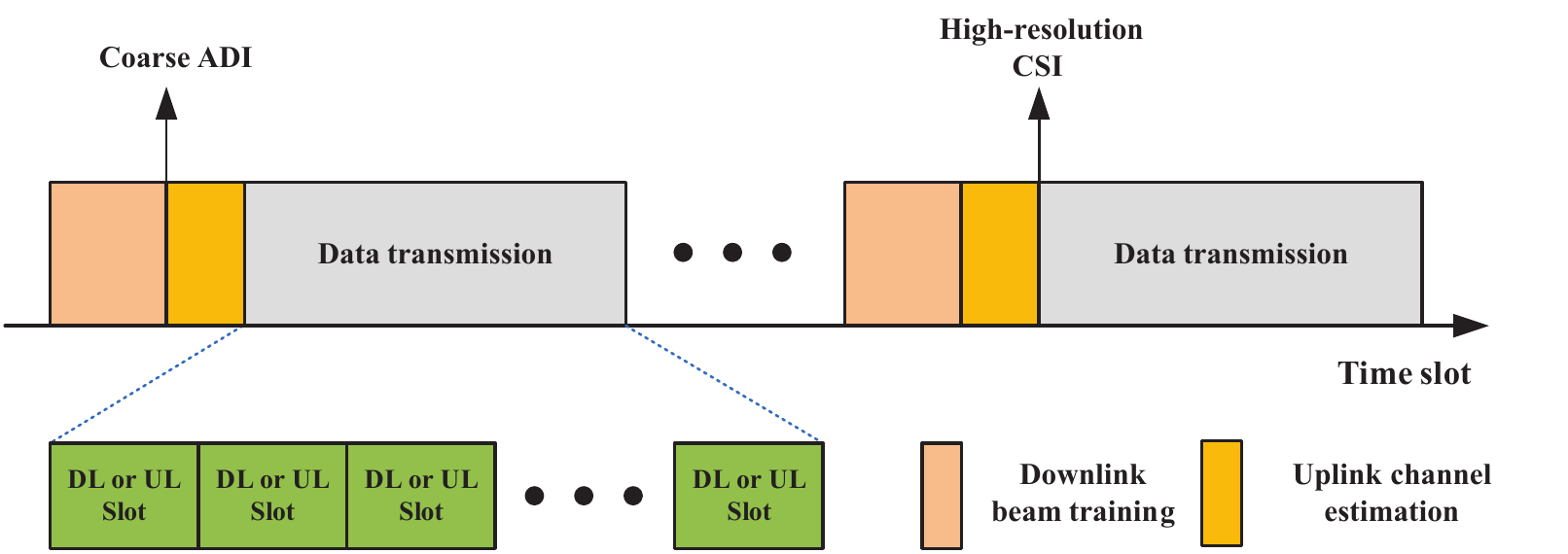}
  \caption{The frame structure of the considered IRS-assisted mmWave system.}
  \label{timeslot}
\end{figure}
\section{Two-Step Cascaded Channel Estimation Protocol}

In this section, we propose a novel two-step cascaded channel estimation protocol to acquire high-resolution CSI, and the specific frame structure is shown in Fig. \ref{timeslot}. First, a downlink beam training procedure is performed to estimate the coarse ADI, and further the rubost UE-IRS-BS communication link is established. In the second step, the coarse ADI is utilized to facilitate the uplink cascaded channel estimation, and meanwhile, the AGMP algorithm is proposed to estimate the high-resolution CSI, which is vitally important to enable further high-quality data transmission. More details will be described in the sequel.

\subsection{First Step: Beam Training based Coarse ADI Acquirement}

\begin{figure}[tp]
  \centering
  \includegraphics[width=0.4\textwidth]{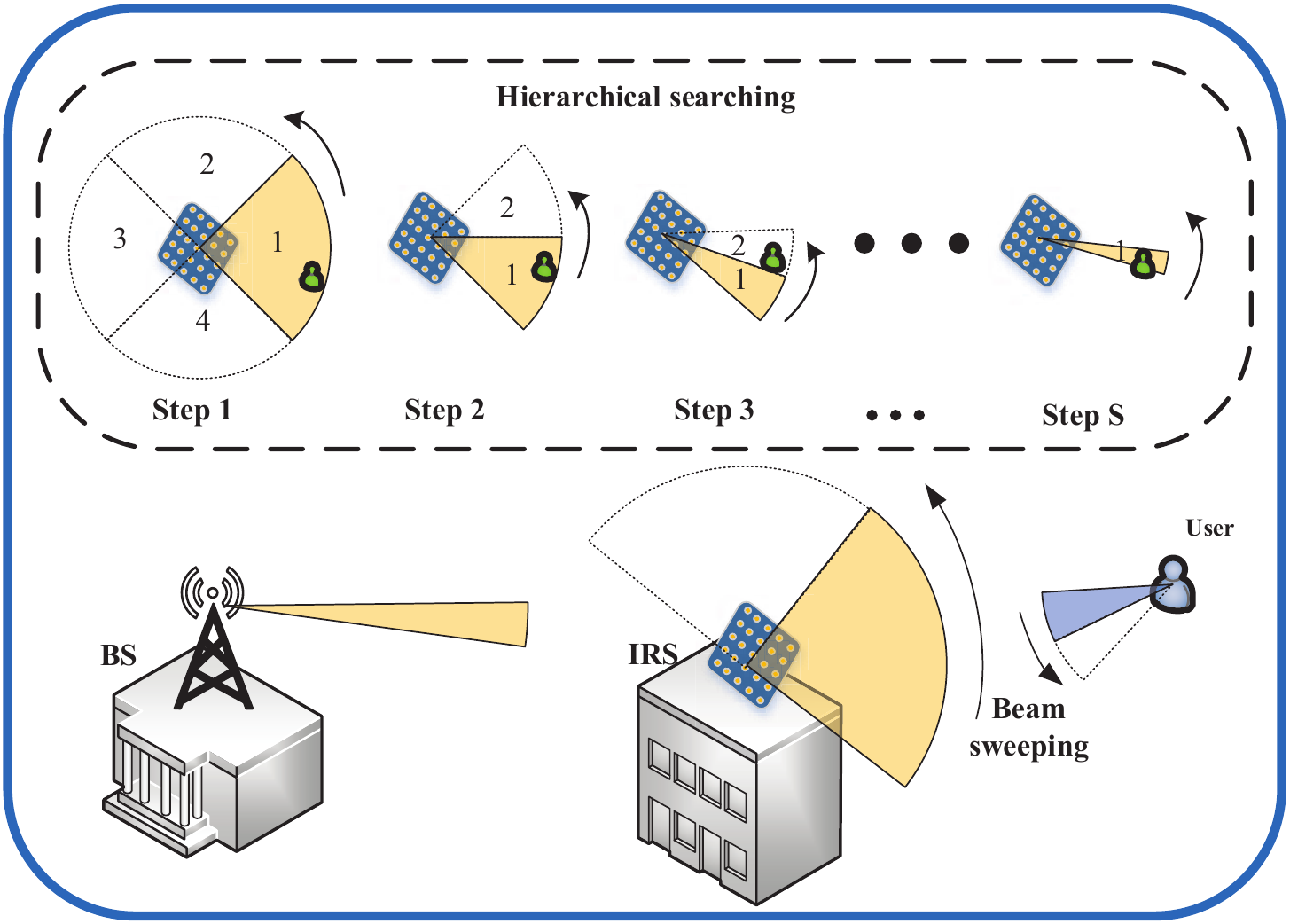}
  \caption{Illustration of the hierarchical beam training protocol for IRS-assisted mmWave MIMO communications.}
  \label{beam_training}
\end{figure}

In the first step, we estimate the coarse ADI of the BS-IRS-UE link by downlink beam training. To do this, we design a hierarchical beam training protocol, which is shown in Fig. \ref{beam_training}. Specifically, we assume that the geometric information of the deployed IRS is \emph{a priori} information, and thus the BS can transmit the training signals towards the IRS directly. Then, the IRS performs hierarchical searching with different beamwidth by adjusting the phase of the passive reflective elements. More specifically, the IRS firstly searches the full-space with wide-beam. At the receiver, the user performs narrow-beam sweeping to determine the optimal beam direction. Subsequently, the wide-beam is further refined until the required phase-resolution of IRS is reached.

In this way, the coarse ADI of both transceivers and IRS can be obtained. However, due to the limited hardware manufacturing capacity, the IRS is fabricated with limited phase-resolution in practice. The performance of beam training based channel estimation can be significantly affected by the ADI estimation error. When the coarse CSI is exploited for further data transmission, the system performance may be degraded severely. In view of this, it is necessary to estimate the high-resolution CSI for achieving the full advantages brought by IRS. Hence, we utilize the coarse ADI to guide the active beamforming at transceivers and passive beamforming at IRS, and thus the uplink can be established to prepare for further uplink cascaded channel estimation, which will be detailed in the next subsection. Furthermore, more details about beam training for IRS-assisted mmWave systems can refer to our prior work \cite{jia2020machine}.

\begin{remark}
  Beam training is an indispensable procedure for initial access of mmWave communications. Here, the coarse ADI obtained by downlink beam training is further utilized to facilitate high-resolution channel estimation with low complexity.

\end{remark}

\subsection{Second Step: Uplink Cascaded Channel Estimation}

\subsubsection{Sparse Problem Formulation}
According to the coarse ADI obtained by the first step, the beamforming vector of user, the combining vector of BS, and the phase-shift matrix of IRS, can be determined, and the robust uplink is further established. Then, the user transmits the uplink pilot signal towards the target IRS with beamforming vector $\mathbf{f}$. Meanwhile, the IRS performs passive beamforming with phase-shift matrix $\mathbf{\Theta}_{s}$ to steer the incident pilot signals towards the BS. Finally, the BS combines the pilot signal with combining vector $\mathbf{w}$. Hence, the received signal at BS is given by
\begin{equation}\label{re_sign}
  y=\mathbf{w}^{H}\mathbf{H}^{B}\mathbf{\Theta}_{s}\mathbf{H}^{I}\mathbf{f}x+\mathbf{w}^{H}\mathbf{n},
\end{equation}
where $x$ is the pilot symbol, and we define $x=1$ for convenience. $\mathbf{n}$ is the Gaussian noise vector with mean zeros and variance $\sigma^{2}$.


Since there are very limited scatters around BS and IRS but rich scatters around the users, we assume that the BS-IRS channel is a rank-one channel, while the UE-IRS channel is regarded as a multi-path channel. According to the channel model in (\ref{channel}), (\ref{re_sign}) can be rewritten as
\begin{equation}\label{refor111}
\begin{aligned}
  y&=\mathbf{w}^{H}\beta\mathbf{a}_{R}\left(\varPhi^{B}\right)\mathbf{a}_{T}\left(\varOmega^{B}\right)^{H}\mathbf{\mathbf{\Theta}}_{s}\mathop{\underset{l=1}{\overset{L}{\sum}}}\alpha_{l}\mathbf{a}_{R}\left(\varPhi_{l}^{I}\right)\mathbf{a}_{T}\left(\varOmega_{l}^{I}\right)^{H}\mathbf{f}\\
  &+\mathbf{w}^{H}\mathbf{n}\\
  &\overset{a}{=}\mathbf{b}_{s}^{H}\overline{\mathbf{H}}\mathbf{f}+\mathbf{w}^{H}\mathbf{n}\\
  &\overset{b}{=}\left(\mathbf{f}^{T}\otimes\mathbf{b}_{s}^{H}\right)\textrm{vec}\left(\overline{\mathbf{H}}\right)+\mathbf{w}^{H}\mathbf{n}\\
  &\overset{c}{=}\left(\mathbf{f}^{T}\otimes\mathbf{b}_{s}^{H}\right)\left(\mathbf{\overline{A}}^{*}_{U}\otimes\mathbf{\overline{A}}_{I}\right)\textrm{vec}\left(\mathbf{\overline{\mathbf{H}}}_{a}\right)+\mathbf{w}^{H}\mathbf{n}\\
  &\overset{d}{=}\left(\mathbf{f}^{T}\otimes\mathbf{b}_{s}^{H}\right)\mathbf{\overline{A}}_{D}\textrm{vec}\left(\mathbf{\overline{\mathbf{H}}}_{a}\right)+\mathbf{w}^{H}\mathbf{n}\\
\end{aligned}
\end{equation}
with
\begin{equation}\label{refor1}
\begin{aligned}
  &\overline{\mathbf{H}}=\mathop{\beta\underset{l=1}{\overset{L}{\sum}}}\alpha_{l}\mathbf{a}_{R}\left(\varPhi_{l}^{I}\right)\mathbf{a}_{T}\left(\varOmega_{l}^{I}\right)^{H},\\
  &\mathbf{b}_{s}=\mathbf{\mathbf{\Theta}}_{s}^{H}\mathbf{a}_{T}\left(\varOmega^{B}\right),
\end{aligned}
\end{equation}
where $\beta$ denotes the channel complex gain of the IRS-BS channel $\mathbf{H}^{B}$, whose AoA and AoD are represented by $\varPhi^{B}$ and $\varOmega^{B}$, respectively. $\overline{\mathbf{H}}\in\mathbb{C}^{N_{I}\times N_{UE}}$ denotes the effective cascaded channel. $\mathbf{b}_{s}\in\mathbb{C}^{N_{I}}$ is the effective combining vector. (a) is derived under the assumption that the position of IRS is \emph{a priori} information at BS, and BS combines the pilot signals from IRS directly, i.e., $\mathbf{w}^{H}\mathbf{a}_{R}\left(\varPhi^{B}\right)=1$. (b) is derived according to the equation $\textrm{vec}\left(\mathbf{ABC}\right)=\left(\mathbf{C}^{T}\otimes\mathbf{A}\right)\textrm{vec}\left(\mathbf{B}\right)$. (c) is derived when the channel matrix is represented in the angular domain, i.e.,
 \begin{equation}
   \overline{\mathbf{H}}=\mathbf{\overline{A}}_{I}\mathbf{\overline{H}}_{a}\mathbf{\overline{A}}_{U}^{H},
 \end{equation}
 where $\overline{\mathbf{A}}_{I}=\left[\mathbf{a}_{R}\left(\overline{\varOmega}_{1}\right),...,\mathbf{a}_{R}\left(\overline{\varOmega}_{G}\right)\right]$ and
$\overline{\mathbf{A}}_{U}=\left[\mathbf{a}_{T}\left(\overline{\varPhi}_{1}\right),...,\mathbf{a}_{T}\left(\overline{\varPhi}_{G}\right)\right]$, which possess the same resolution $G$ and constitute the dictionary matrix $\mathbf{\overline{A}}_{D}=\mathbf{\overline{A}}^{*}_{U}\otimes\mathbf{\overline{A}}_{I}\in\mathbb{C}^{N_{UE}N_{I}\times G^2}$, where each column has the form $\mathbf{a}_{T}\left(\overline{\varPhi}_{g}\right)^{*}\otimes\mathbf{a}_{R}\left(\overline{\varOmega}_{g}\right), g=1,...,G$. $\mathbf{\overline{\mathbf{H}}}_{a}$ is a sparse angular domain channel matrix consisting of only $L$ non-zero elements. The AoAs/AoD pairs of different paths can be determined according to the positions of the non-zero elements in $\mathbf{\overline{\mathbf{H}}}_{a}$, and the values of these elements are the corresponding channel complex gains.

In this way, the cascaded channel estimation is converted into a sparse recovery problem, which can be solved by basic pursuit theory \cite{chen2001atomic}. More details will be described in the sequel.


\subsubsection{AGMP Solution}

  Regarding the coarse ADI obtained by beam training as prior information, a low-complexity AGMP algorithm is designed to estimate the high-resolution cascaded CSI in this subsection. In particular, following reference \cite{7454701}, we only estimate LOS path while neglecting other non-LOS (NLOS) paths. The reason is that the energy of LOS path is dozens of times stronger than that of NLOS paths according to practical field measurements  \cite{6834753}, and thus the influence of NLOS paths is dispensable for mmWave communications. In particular, our AGMP algorithm can be readily extended to the multi-path channel estimation cases by estimating the paths one by one \cite{6847111}. More details about the proposed AGMP algorithm can refer to \textbf{Algorithm \ref{ACS}}.

\begin{algorithm}[t]
  \caption{AGMP Based High-resolution Channel Estimation}
  \label{ACS}
  \begin{algorithmic}[1]
  \INPUT Coarse ADI estimated by beam training: $\widehat{\varOmega}$ and $\widehat{\varPhi}$ for user and IRS, respectively.

  Parameters: the resolution of the dictionary matrix $\widetilde{G}$, the number of iteration $\zeta$, phase-resolution of IRS $R_{I}$, phase-resolution of user $R_{U}$.

  Received signal at BS $y$.

  \OUTPUT The effective cascaded channel $\widehat{\mathbf{H}}$.

  \textbf{Step 1}: \textit{Generate adaptive grid}
  \STATE Determine the range of ADI estimation error $c_{1}=2\pi/R_{U}$ for user and $c_{2}=2\pi/R_{I}$ for IRS.
  \STATE Generate the dictionary matrix $\widetilde{\mathbf{A}}_{D}$ by dividing the coarse angular range into $\widetilde{G}$ grids:
  \begin{equation}
  \begin{aligned}
    &\widetilde{\mathbf{A}}_{U}=\left[...,\mathbf{a}_{T}\left(\widehat{\varOmega}-c_{1}/2+\frac{c_{1}}{\widetilde{G}}g\right),...\right],\\
    &\widetilde{\mathbf{A}}_{I}=\left[...,\mathbf{a}_{R}\left(\widehat{\varPhi}-c_{2}/2+\frac{c_{2}}{\widetilde{G}}g\right),...\right],\\
    &g=0,...,\widetilde{G}-1,\\
    &\widetilde{\mathbf{A}}_{D}=\widetilde{\mathbf{A}}_{U}^{*}\otimes\widetilde{\mathbf{A}}_{I}.
  \end{aligned}
  \end{equation}
  \STATE Generate beamforming vector $\mathbf{f}$ of user and efficient combining vector $\mathbf{b}_{s}$ according to the coarse ADI.
  \STATE Generate the sensing matrix
  \begin{equation}
  \widetilde{\mathbf{Q}}=\left(\mathbf{f}^{T}\otimes\mathbf{b}_{s}^{H}\right)\mathbf{\widetilde{A}}_{D}.
  \end{equation}

  \textbf{Step 2}: \textit{Matching pursuit based high-resolution cascaded channel estimation}

  \STATE $\mathcal{I}_{0}=$ empty set, residual $r_{0}=y$, set the iteration counter $t=1$.
  \FOR{$t \leqslant \zeta$}
  \STATE $g=\arg\max\left|\widetilde{\mathbf{Q}}\left(g\right)^{H}r_{t-1}\right|,g=1,...,\widetilde{G}^{2}$
  \STATE $\mathcal{I}_{t}=\mathcal{I}_{t-1}\cup \left\{ g\right\}$
  \STATE $\mathbf{h}_{t}=\arg\min\left\Vert y-\widetilde{\mathbf{Q}}_{\mathcal{I}_{t}}\mathbf{h}_{t}\right\Vert $
  \STATE $r_t=y-\widetilde{\mathbf{Q}}_{\mathcal{I}_{t}}\mathbf{h}_{t}$
  \STATE $t=t+1$
  \ENDFOR
  \STATE For $g \in \mathcal{I}_{t-1}$, $\widehat{\mathbf{h}}_{a}\left(g\right)=\mathbf{h}_{t}\left(g\right)$ and $\widehat{\mathbf{h}}_{a}\left(g\right)=0$ otherwise.
  \STATE $\widehat{\mathbf{H}}_{a}=\textrm{unvec}\left(\widehat{\mathbf{h}}_{a}\right)$.
  \RETURN $\widehat{\mathbf{H}}=\widetilde{\mathbf{A}}_{I}\widehat{\mathbf{H}}_{a}\widetilde{\mathbf{A}}_{U}^{H}$.

  \end{algorithmic}
\end{algorithm}

The operations of \textbf{Algorithm \ref{ACS}} are summarized as follows. In step 1-4, the dictionary matrix and sensing matrix, also known as grid, are generated adaptively according to the coarse ADI obtained by the first step. Then, matching pursuit algorithm is utilized to estimate high-resolution CSI. For each iteration, the column of $\widetilde{\mathbf{Q}}$, which is most strongly correlated with residual, is selected (step 7) to update the column index set (step 8), in which each element is related to an AoA/AoD pair from the quantized grid $\widetilde{\mathbf{A}}_{U}$ and $\widetilde{\mathbf{A}}_{D}$. Then, the channel gains corresponding to selected AoA/AoD pairs are obtained by step 9, and the residual is updated by removing the contribution of the selected columns (step 10). When the maximal number of iterations is reached, the angular domain channel $\widehat{\mathbf{H}}_{a}$ is estimated (step 13-14). Therefore, the estimated effective cascaded channel $\widehat{\mathbf{H}}$ can be constructed with high-resolution (step 15).

\textbf{Complexity analysis}: For standard CS theory based mmWave channel estimation, where random beamforming is performed without the assistance of coarse ADI, the computation complexity is $\mathcal{O}\left(\zeta N_{T}N_{R}G^{2}\right)$ ($G\geqslant \max \left(N_T,N_R\right)$), which is proportional to dimension of the dictionary matrix $G^2$. Moreover, it requires at least $N_{T}N_{R}$ measurements to recover the sparse channel. Due to the large number of antennas equipped at both transceivers and IRS for IRS-assisted mmWave MIMO communications, the computation complexity is almost prohibitive. For the proposed AGMP algorithm, by exploiting the coarse ADI obtained by beam training, the dimension of dictionary matrix can be significantly reduced $\widetilde{G}\ll\min \left(N_T,N_R\right)$, which further reduces the computation complexity of channel estimation to $\mathcal{O}\left(\zeta N_{T}N_{R}\widetilde{G}^{2}\right)$. In particular, the simulation results show that the SE performance can converge to the upper bound even when $\widetilde{G}=3$.

\begin{remark}
 The performance of the benchmark scheme can be severely impaired by imperfect hardware manufacturing technique. To address this challenge, the coarse ADI obtained by beam training in the first step is fully utilized to reduce the dimension of the dictionary matrix, which enables the high-resolution channel estimation with simplified computation complexity of its solution.
 \end{remark}

\section{Simulation Results}

%
%

In this section, the performance of the proposed two-step cascaded channel estimation protocol is evaluated via computer simulation. In our simulation, the corresponding system parameters are set as following: the number of antennas at BS $N_{BS}=64$, the number of antennas at IRS $N_{I}=64$, the number of antennas at user $N_{UE}=16$, the carrier frequency $f_c=28$ GHz, the number of paths from IRS to user $L=3$.

\begin{figure}[tp]
  \centering
  \includegraphics[width=0.4\textwidth]{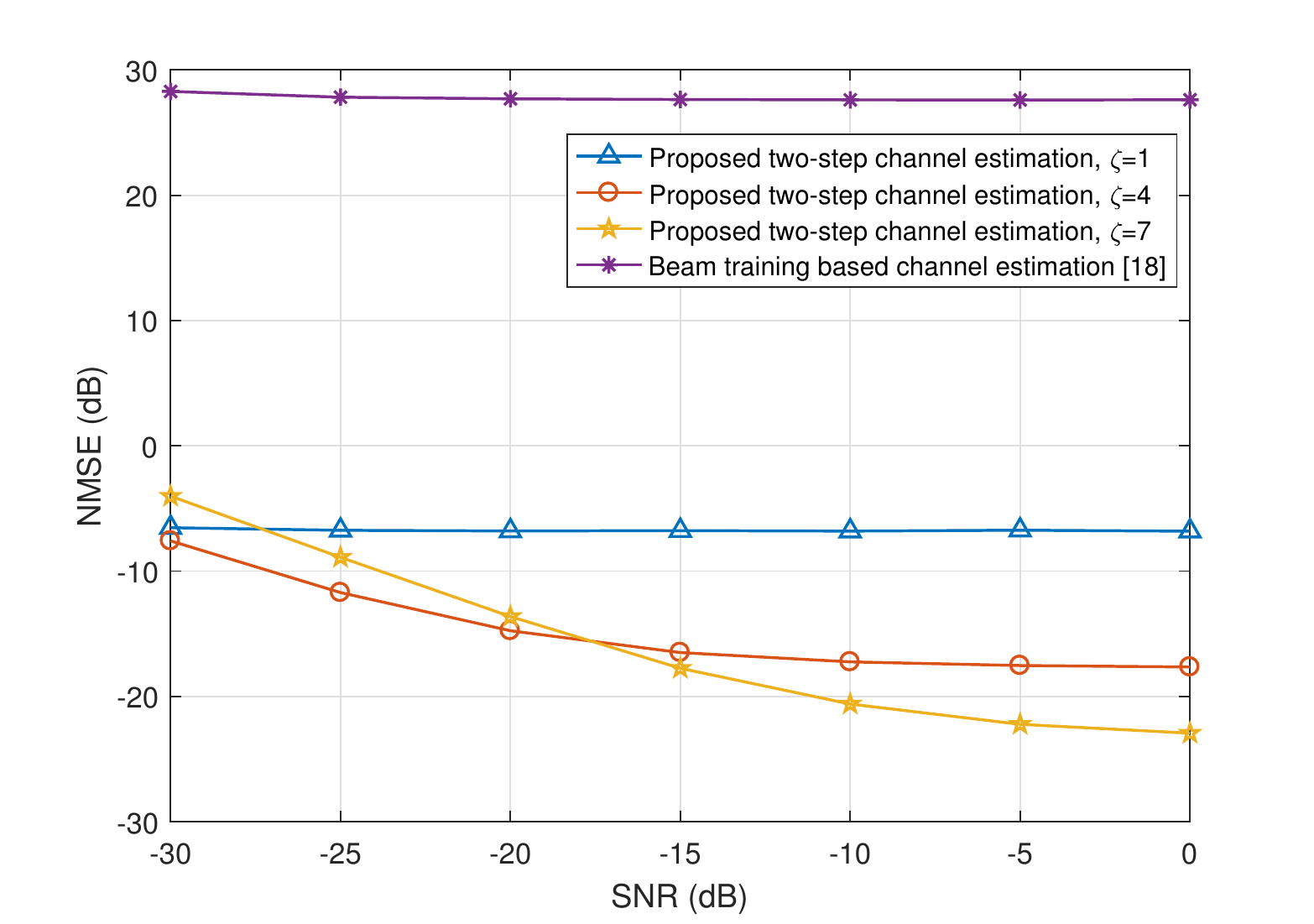}
  \caption{The NMSE performance comparison of the benchmark scheme and the two-step channel estimation protocol under different iterations, the dictionary resolution $\widetilde{G}=5$.}
  \label{NMSE_SNR}
\end{figure}
 We evaluate the normalized mean square error (NMSE) performance of our proposed two-step concatenated channel estimation protocol against signal-to-noise ratio (SNR). As shown in Fig. \ref{NMSE_SNR}, assuming the dictionary resolution $\widetilde{G}=5$, the NMSE performance against SNR is compared under different iterations of the AMGP algorithm. The NMSE is defined as
\begin{equation}
 \text{NMSE}=10\log_{10}\left(\mathbb{E}\left[\left\Vert \mathbf{H}-\widehat{\mathbf{H}}\right\Vert _{F}^{2}/\left\Vert \mathbf{H}\right\Vert _{F}^{2}\right]\right),
\end{equation}
where $\mathbf{H}$ is the perfect CSI, and $\widehat{\mathbf{H}}$ is the estimated CSI. Moreover, we assume that there are a single LOS path and two NLOS paths between the IRS and user, and the Rician K-factor is set as 20dB. Obviously, even with few iterations, the proposed scheme shows great superiority over the benchmark scheme \cite{ning2019channel}, in which the CSI is obtained directly by beam training. Moreover, we can observe that increasing the iteration will improve the NMSE performance. In particular, the beam training based channel estimation scheme suffers from poor NMSE performance even at high-SNR, and thus we can conclude that the ADI estimation plays a dominant role for the accuracy of channel estimation, especially for high-directional mmWave channel.
\begin{figure}[tp]
  \centering
  \includegraphics[width=0.4\textwidth]{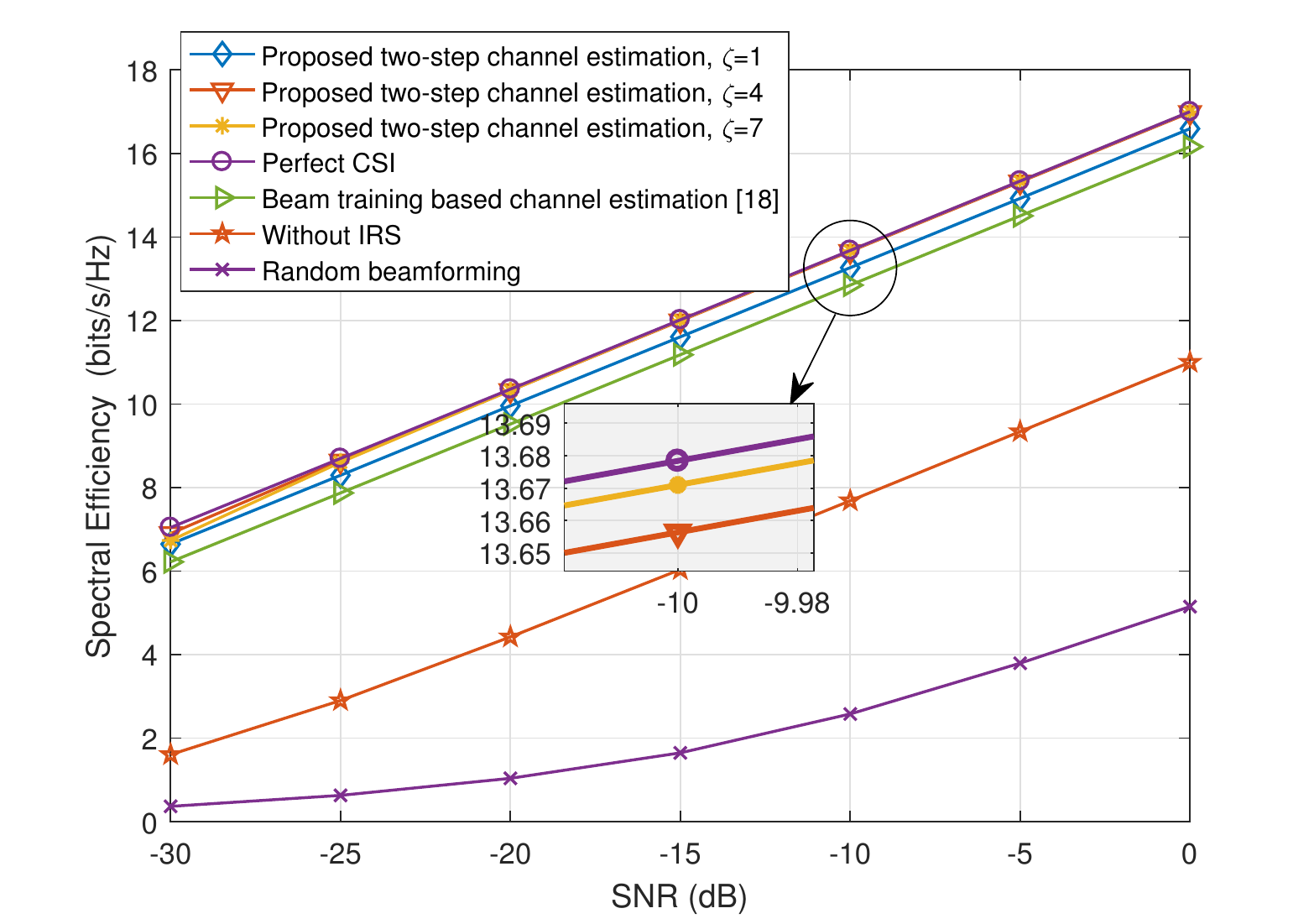}
  \caption{The SE performance comparison of the benchmark scheme and the two-step channel estimation protocol under different iterations, the dictionary resolution $\widetilde{G}=5$.}
  \label{SE_SNR}
\end{figure}

Fig. \ref{SE_SNR} shows the comparison of SE performance against SNR. In particular, the SE performance achieved by perfect CSI is adopted as the upper bound, and the random beamforming is shown as the lower bound. We can observe that our proposed two-step channel estimation protocol outperforms the benchmark scheme even with limited dictionary resolution ($\widetilde{G}=5$) and iteration $\zeta=4$, and meanwhile can achieve the near-optimal SE performance obtained by perfect CSI with the increase of iteration. Moreover, we also present the SE performance for the case without IRS, which verifies that the presence of IRS indeed improves the SE performance of mmWave communications due to the additional beamforming gain brought by IRS. Comparing Fig. \ref{NMSE_SNR} with Fig. \ref{SE_SNR}, we can conclude that the beam training based channel estimation scheme can only estimate coarse ADI due to the limited phase-resolution of IRS, while the estimated channel gain is not accurate drastically due to the effects of the angular estimation error, multi-path and noise. When the coarse CSI is adopted for data detection or system optimization, poor performance may be generated. In this regard, we propose a high-resolution channel estimation scheme to mitigate the influence of these factors by exploiting an AGMP algorithm. Since the IRS can reinforce the signal power through directional beamforming, which is equivalent to degrade the influence of the noise, the proposed protocol can still reap a great NMSE performance even at low-SNR for IRS-assisted mmWave MIMO communications.

%

\begin{figure}[tp]
  \centering
  \includegraphics[width=0.4\textwidth]{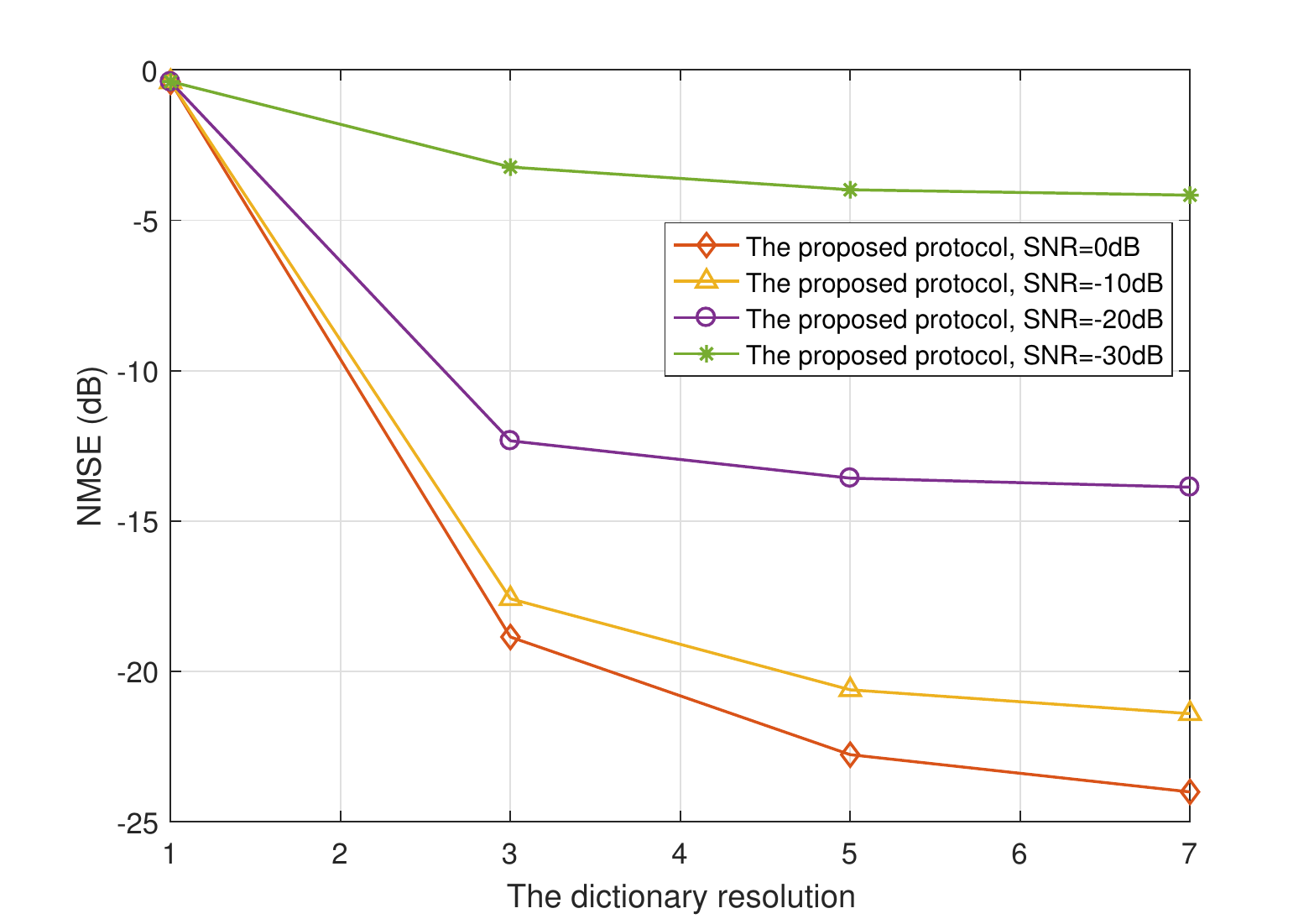}
  \caption{The NMSE performance against the dictionary resolution $\widetilde{G}$ under different SNR cases, $\zeta=7$.}
  \label{NMSE_G_Z7}
\end{figure}

\begin{figure}[tp]
  \centering
  \includegraphics[width=0.4\textwidth]{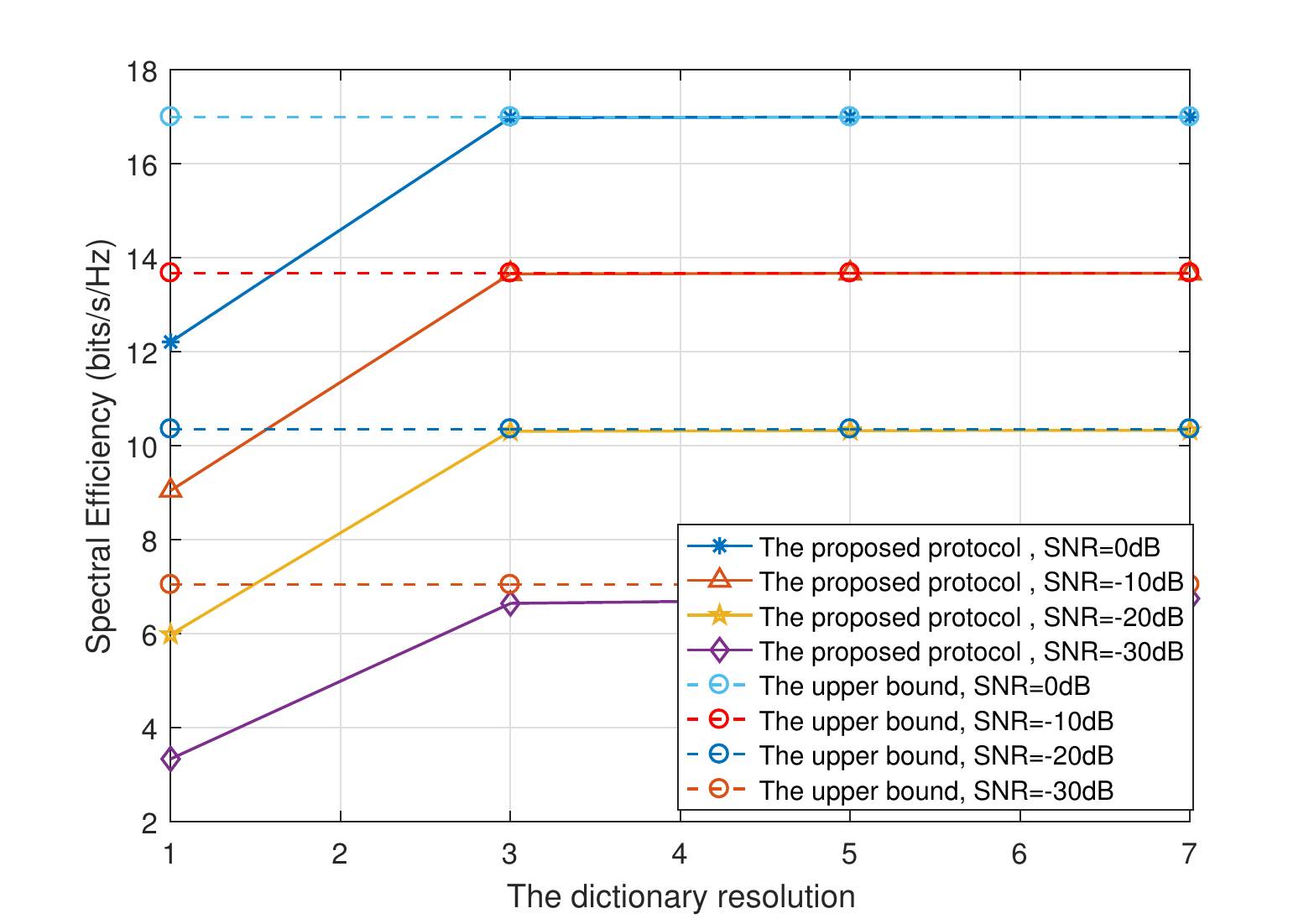}
  \caption{The SE performance against the dictionary resolution $\widetilde{G}$ under different SNR cases, $\zeta=7$.}
  \label{SE_G_Z7}
\end{figure}


Fig. \ref{NMSE_G_Z7} and Fig. \ref{SE_G_Z7} compare the NMSE performance and SE performance against the dictionary resolution $\widetilde{G}$ under different SNR cases, respectively. With the increase of the dictionary resolution $\widetilde{G}$, the NMSE and SE performance can be significantly improved for different SNR cases. In particular, the SE performance converges to the upper bound obtained by perfect CSI quickly even when $\widetilde{G}=3$, which suggests that $\widetilde{G}=3$ is the optimal selection for the given system settings.
\section{Conclusion}
In this paper, we have studied the high-resolution cascaded channel estimation issue for IRS-assisted mmWave MIMO communications. Specifically, by exploiting the sparsity of mmWave channels, we formulate the cascaded channel estimation problem into a sparse recovery problem, and propose a two-step cascaded channel estimation protocol to achieve high-resolution channel estimation. The simulation results verify the advantages of our proposed protocol in terms of SE and NMSE performance as compared to the beam training based cascaded channel estimation scheme. Moreover, our proposed AGMP algorithm can approach the optimal performance achieved by perfect CSI with low computation complexity.


%

%

\ifCLASSOPTIONcaptionsoff
  \newpage
\fi



%

\bibliographystyle{IEEEtran}
\bibliography{CE_PIMRC}

\end{document}